\documentclass[aps,amsmath,amssymb,reprint]{revtex4-2}
\usepackage{graphicx}
\begin{document}

\title{Unconventional superconductivity in Sc$_2$Ir$_{4-x}$Si$_x$ by spin-orbit coupling driven flat band}
\author{Zhengyan Zhu$^{\star}$, Yuxiang Wu$^{\star}$, Shengtai Fan, Yiliang Fan, Yiwen Li, Yongze Ye, Xiyu Zhu$^{\dag}$, Haijun Zhang, Hai-Hu Wen$^{\ast}$}

\affiliation{Center for Superconducting Physics and Materials, National Laboratory of Solid State Microstructures and Department of Physics,  Collaborative Innovation Center for Advanced Microstructures, Nanjing University, Nanjing 210093, China}

\date{\today}

\begin{abstract}
We have successfully synthesized the Laves phase superconductors with Si doping to the Ir sites in Sc$_2$Ir$_{4-x}$Si$_{x}$ with a kagome lattice. The nonmonotonic and two-dome-like doping dependence of the superconducting transition temperature $T_{\rm c}$ was observed. For some samples, especially Sc$_2$Ir$_{3.5}$Si$_{0.5}$ with the optimal $T_{\rm c}$, they exhibit non-Fermi liquid behavior at low temperatures, including the divergence of the specific heat coefficient and the semiconducting-like resistivity. Around the optimal doping, it shows strong superconducting fluctuations by the resistivity and the possible $d$-wave pairing by the superconductivity related specific heat. Combined with the first-principles calculations, these strongly suggest unconventional superconductivity and correlation effect in this system, which is mainly induced by the flat band effect when considering the strong spin-orbit coupling (SOC).
\end{abstract}

\maketitle

The kagome lattices have drawn great attention in the community due to their inherent geometrical spin frustration \cite{1}. They provide good platforms for exploring many exotic electronic states such as quantum spin liquid \cite{2,3,4} and other intriguing charge or magnetic orders \cite{5,6}. Subsequently, more focus has been directed towards the electronic band structure of the kagome lattices, which is expected to possess saddle points, topological Dirac cones, and flat bands \cite{7,8,9}. Consequently, superconductors with kagome lattices may have potential unconventional pairing mechanisms \cite{10}. Notably, the Laves phases, as a well-known family of intermetallic compounds with kagome lattices, have diverse crystal structures such as the cubic $C15$-type (MgCu$_2$) and the hexagonal $C14$-type (MgZn$_2$) \cite{11,12}. Several $C15$-type Laves phase superconductors, such as CeRu$_2$ and ZrV$_2$ \cite{13,14}, have been reported as possible unconventional superconductors. Recently, due to the strong SOC of Ir, researches have been focused on the Iridium-based $C15$ Laves phases $A$Ir$_2$ ($A$ = Ca, Sr, Ba, Th, Sc, Zr) \cite{18,19,20,21,22}. Many reports suggest the possibility of anomalous electronic states and possible unconventional superconductivity \cite{23,24,25}. Among them, the $T_{\rm c}$ of ScIr$_2$ was reported to be 2 $\sim$ 2.4 K long time ago \cite{33}. Early calculations of the band structures for ScIr$_2$ have confirmed that the density of states (DOS) near the Fermi level is mainly contributed by the 5$d$ electrons of Ir, and the band structure is highly influenced by the strong SOC effect \cite{34,35}.

Some special energy bands without or with weak dispersions are called as flat bands, which themselves have fascinating properties, such as the correlated insulating states and unconventional superconductivity \cite{definefb}. The flat bands have a large effective mass and relatively small kinetic energy of electrons. As a result, many body effects dominate, leading to the strong correlation effects and non-Fermi-liquid behavior \cite{flatband}. Clear evidence of flat bands has been observed in the kagome materials such as $A$V$_3$Sb$_5$ and $A$Ti$_3$Bi$_5$ ($A$ = K, Rb, and Cs) associated with unconventional superconductivity \cite{CsVSbflat,CsTiBiflat}. In addition, when considering SOC, the electronic bands tend to split, which sometimes also produce flat bands \cite{SOCfb}. Many theoretical and experimental studies were initiated on the relationship between flat bands and superconductivity \cite{64,66}. For kagome superconductor LaRu$_3$Si$_5$, the presence of flat bands brings about the high DOS at the Fermi energy, which enhances its $T_{\rm c}$ \cite{LaRuSi}. In addition to enhanced superconducting pairing, flat bands also lead to strong superconducting fluctuations (SCFs) \cite{65}. In addition, exotic quantum phenomena are often observed in materials with flat bands due to the induced strong correlation effect. For the well-known twist graphene with magic angles, the flat band at half-filling leads to the correlated insulating states \cite{Mag}. For kagome metal Ni$_3$In, the strong electron-electron interactions induced by their flat bands lead to the upturn of the specific heat coefficient observed at low temperatures \cite{NiIn}. Superconductors with electronic flat bands deserve more in-depth studies, including their exotic physical properties and the potential unconventional superconductivity.

In this study, a nonmonotonic and two-dome like doping dependence of $T_{\rm c}$ was observed for a series of Sc$_2$Ir$_{4-x}$Si$_x$ ($x=0\sim0.7$). For some samples in the second dome region, we observed non-Fermi-liquid behavior at low temperatures in their normal states, which exhibit as a semiconducting ground state and a divergence of the specific heat coefficient. At the optimal doping point, the superconductivity is enhanced, showing pronounced SCFs. Furthermore, the normal-state specific heat data deviates from the Debye model and the superconducting-state specific heat exhibits possible $d$-wave pairing. All these can be attributed to the strong correlation effect, which is mainly induced by the tunable flat band near the Fermi surface, relevant to the kagome structure and the SOC effect.

\begin{figure}
  \includegraphics[width=8.8 cm]{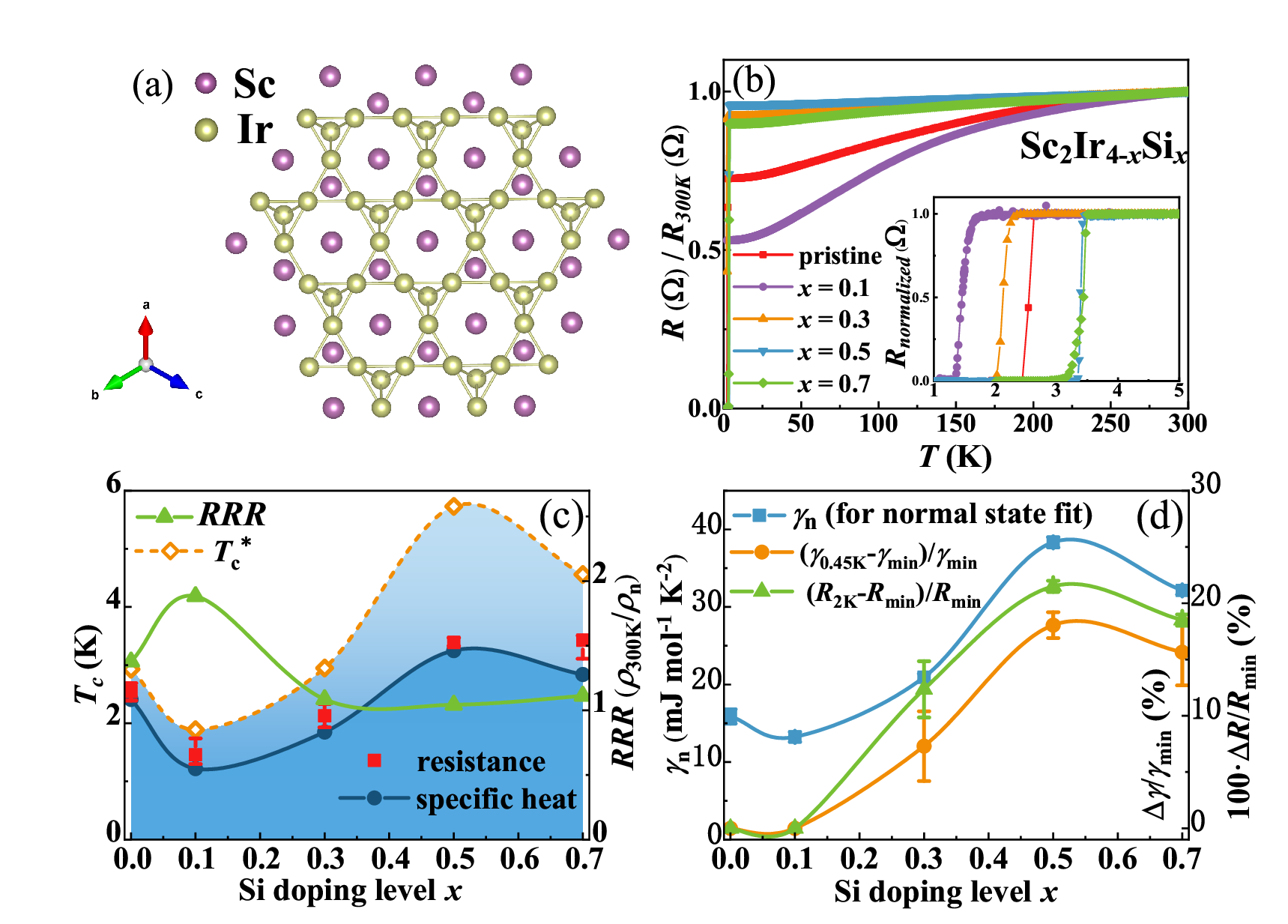}
\caption{(a) The kagome lattice of Ir for ScIr$_2$. (b) The normalized resistance of Sc$_2$Ir$_{4-x}$Si$_x$ from 2 K to 300 K. The inset shows the enlarged view of the superconducting transitions around $T_{\rm c}$. (c) The phase diagram of Sc$_2$Ir$_{4-x}$Si$_x$ involving both $T_{\rm c}$ (measured by resistivity and specific heat) and $RRR$. (d) Doping dependence of the electronic specific heat coefficient $\gamma_n$ in the normal state, the degree of the increase of resistivity measured by [$R_{\rm 2 K}-R_{\rm min}$]/$R_{\rm min}$, and the divergence of specific heat coefficient measured by [$\gamma_{\rm 0.45K}-\gamma_{\rm min}$]/$\gamma_{\rm min}$ for these samples. Here, both $\gamma_{\rm min}$ and $R_{\rm min}$ are the associated values at the minimum points of resistivity and specific heat coefficient in the normal state, respectively.} \label{fig1}
\end{figure}

For the crystal structure of ScIr$_2$, Ir atoms form kagome lattices in the planes perpendicular to the [111] direction shown in Fig.~\ref{fig1}(a). We synthesized a series of compounds of Sc$_2$Ir$_{4-x}$Si$_x$ ($x=0\sim0.7$) by the arc melting method \cite{44}. Similar to several other ternary $C15$-type Laves phases reported previously (e.g., Mg$_2$Ni$_{3.2}$Ge$_{0.8}$, Mn$_2$Ni$_3$Ge, Nd$_2$Ni$_{3.5}$Si$_{0.5}$ and Mg$_2$Ni$_3$As) \cite{15,16,17}, Ir atoms are partially substituted by nonmetallic Si atoms, which are randomly distributed in the kagome lattice. The high quality and bulk superconductivity of all samples were confirmed by the Powder x-ray diffraction and magnetic susceptibility measurements (see Supplementary Materials) \cite{44}. Then we performed further studies. The resistance of all samples are shown in Fig.~\ref{fig1}(b), one can see that the variation of $T_{\rm c}$ is nonmonotonic with increasing doping level of Si. We summarized their $T_{\rm c}$ and residual resistance ratio ($RRR$ = $R_{\rm 300K}/R_{\rm 5K}$), plotted them in a phase diagram in Fig.~\ref{fig1}(c), which reveals two superconducting domes. Interestingly, the doping dependence of $T_{\rm c}$ is opposite to the $RRR$. The fact that the values of $RRR$ in highly doped samples are close to 1 suggests the disordered distribution of Sc atoms in the lattice, which is similar to the high-entropy alloys and other doped metals \cite{HEA}. Initially, the introduction of Si atoms enhances the impurity scattering and destroys the superconductivity \cite{45}. But the $RRR$ increases with a light doping. Then, the $T_{\rm c}$ value increases again with further doping Si, while the $RRR$ decreases and approaches to 1. The highest $T_{\rm c}$ is then observed at the doping level around $x=0.5$ with the lowest $RRR$ about 1.04. The presence of nonmonotonic doping or pressure dependent $T_{\rm c}$ in the phase diagram in many systems is quite often accompanied by the occurrence of unconventional superconductivity, such as in cuprates, iron-based superconductors, and recently discovered CsV$_3$Sb$_5$ \cite{46,47,48,49,cuprates}. The origin of nonmonotonic $T_{\rm c}$ dependence remains elusive and deserves further study.

\begin{figure}
  \includegraphics[width=8.9 cm]{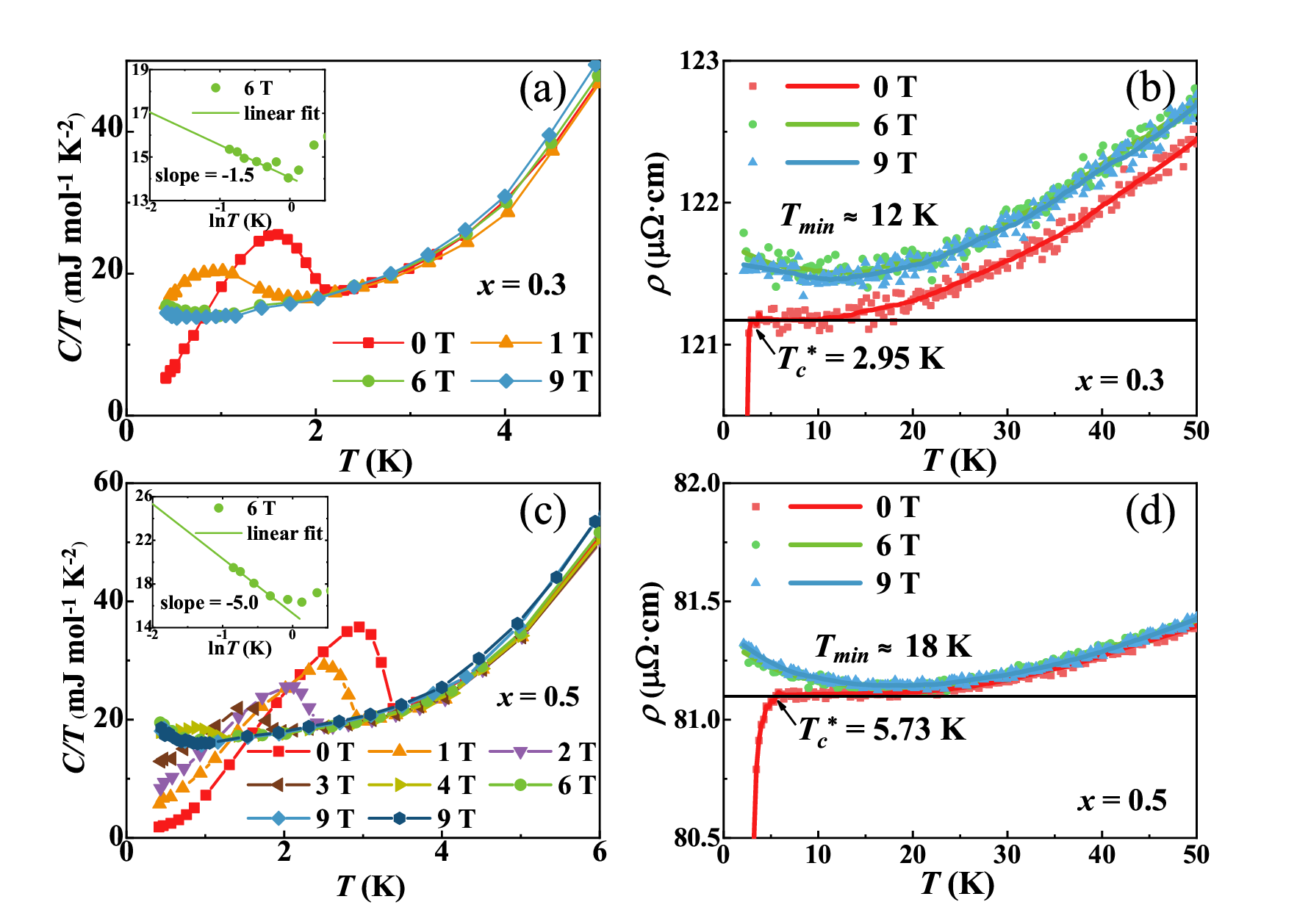}
\caption{Temperature dependence of specific heat under different magnetic fields below 6 K for (a) Sc$_2$Ir$_{3.7}$Si$_{0.3}$, (c) Sc$_2$Ir$_{3.5}$Si$_{0.5}$. Insets show the enlarged view of the linear dependence of $C/T$ on ln$T$. Temperature dependence of resistivity under various magnetic fields from 2 K to 50 K for (b) Sc$_2$Ir$_{3.7}$Si$_{0.3}$, (d) Sc$_2$Ir$_{3.5}$Si$_{0.5}$.} \label{fig2}
\end{figure}

Figure~\ref{fig2}(a) and ~\ref{fig2}(c) show the low temperature specific heat of Sc$_2$Ir$_{3.7}$Si$_{0.3}$ and Sc$_2$Ir$_{3.5}$Si$_{0.5}$. When the magnetic field completely suppresses superconductivity, the specific heat coefficient increases slightly at very low temperatures for Sc$_2$Ir$_{3.7}$Si$_{0.3}$. And for Sc$_2$Ir$_{3.5}$Si$_{0.5}$ with higher $T_{\rm c}$, the increasing trend is more pronounced at low temperatures, clearly indicating the divergence of the specific heat coefficient. Such upturns emerge only after superconductivity is completely suppressed by the magnetic field. We can rule out the possibility of the Schottky anomaly in specific heat \cite{50}. Because if it is the case, the divergence (as part of the peak of Schottky anomaly) at zero and low fields, should gradually become broadened and finally smeared out at higher fields \cite{51,52}. In fact, however, the data at 6 and 9 T overlap nicely. Therefore, the observed upturn of the specific heat coefficient at low temperatures should be the intrinsic property of the normal state. As shown in the insets of Fig.~\ref{fig2}(a) and ~\ref{fig2}(c), the specific heat coefficient $C/T$ at low temperatures is well fitted by the linear relationship with ln$T$. This gives the signature of quantum fluctuations for strange metals, whose electronic specific heat shows the logarithmic temperature dependence in the low temperature limit $C_e /T= A$ln$(T_0/T)$ around the quantum critical point, sometimes accompanied by the linear $T$ dependent resistivity \cite{53,54,55,CeRhGe}. And in the heavy fermion superconductors, the upturn of specific heat coefficient is attributed to the enormous enhancement of the effective mass of the quasiparticles, due to the strong correlation effect \cite{HF}. Fig.~\ref{fig1}(d) shows the evolution of $\gamma_n$ with doping. For Sc$_2$Ir$_4$ and Sc$_2$Ir$_{3.9}$Si$_{0.1}$, we get $\gamma_n$ by a linear extension of the normal-state specific heat coefficient to 0 K. And for other samples with the low-temperature divergence, we obtain $\gamma_n$ by fitting the data to the logarithmic relation. $\gamma_n$ represents the DOS in the normal state near the Fermi energy. Considering the unphysical infinitely large value of $\gamma_n$ given by that logarithmic equation, we choose the $\gamma_n$ at 0.01 K. It shows a strong increase in the second superconducting dome, especially at the optimal doping point with the highest $T_{\rm c}$. Therefore, the logarithmic increase of specific heat coefficient observed here indicates the divergence of the DOS and the electronic effective mass, which is related to the enhancement of the electronic correlation and quantum fluctuations.

\begin{figure}
  \includegraphics[width=9 cm]{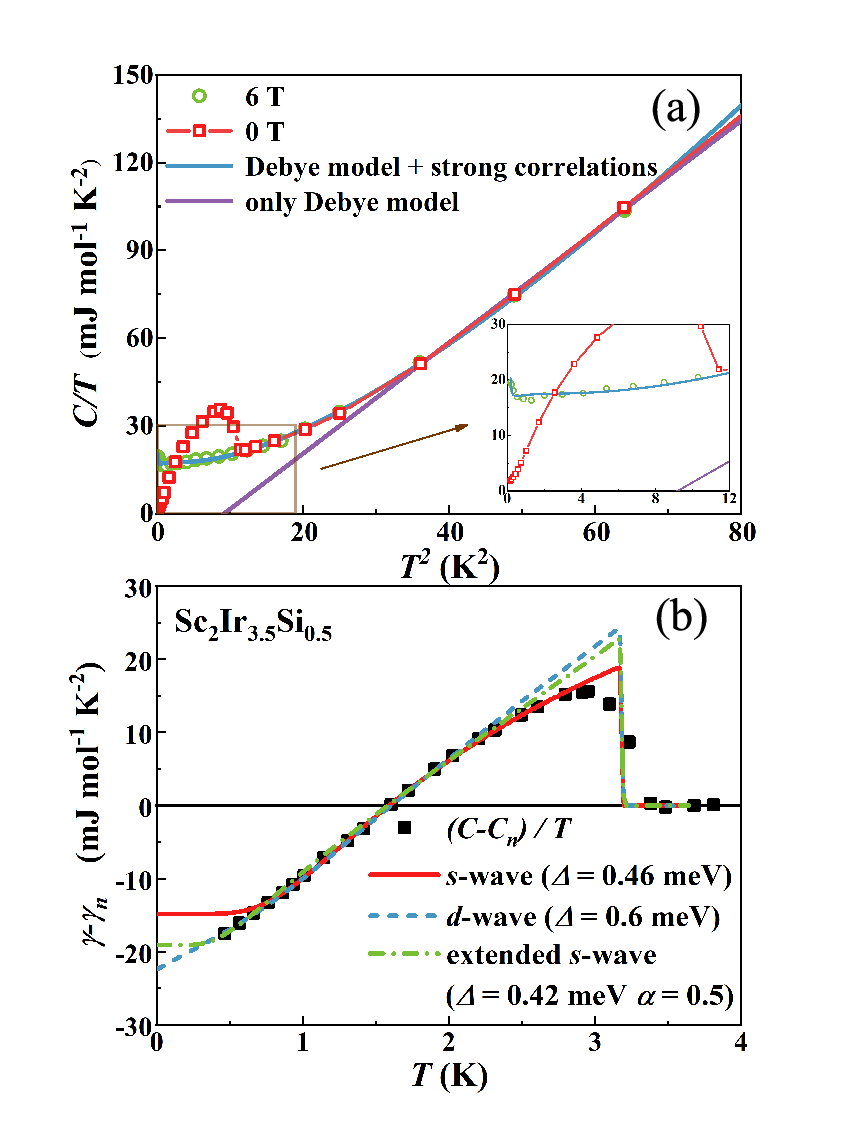}
\caption{(a) The specific heat coefficient $C/T$ versus $T^2$ at 0 T and 6 T. The normal state data (as green symbols) deviates from the Debye model (as purple line) but can be well fitted when considering the electron correlations (as blue line). The inset is an enlarged view of the fits at low temperatures. (b) The electronic specific heat coefficient at 0 T and the fitting curves with three different gap structures.} \label{fig3}
\end{figure}

As shown in Fig.~\ref{fig2}(b) and ~\ref{fig2}(d), the electric transport properties also show the anomalous ground state. For Sc$_2$Ir$_{3.7}$Si$_{0.3}$, the resistivity increases slightly below 12 K at 6 and 9 T, behaving like a semiconductor. The semiconducting behavior is more pronounced for Sc$_2$Ir$_{3.5}$Si$_{0.5}$. Both samples behave as the Fermi liquid above $T_{\rm c}$ at 0 T. This is diametrically opposed to the well-known Kondo effect \cite{58}, for which, the resistivity increases below the Kondo temperature at 0 T due to the localized magnetic moments of the impurities, and eventually behaves as the Fermi liquid when the magnetic field is sufficiently high \cite{59}. Moreover, the itinerant electrons, rather than localized electrons caused by some defects or disorders, form paramagnetic ground state with very small local magnetic moments and magnetoresistance \cite{44}. Therefore, the weak Anderson localization should also be excluded, since it mostly occurs in disordered systems and exhibits the metal-insulator transition which may be suppressed by magnetic fields \cite{60,61}. For some underdoped cuprates, people observed that the resistivity shows an upturn at low temperatures with the same logarithmic relation when superconductivity is suppressed, and this upturn is supposed to originate from large Hubbard interactions induced by strong correlation effects \cite{38,39}. Furthermore, the relatively strong SOC may lead to nontrivial band topologies such as flat bands, thus realizing exotic quantum states and sometimes inducing unconventional insulating state \cite{29,30}. In present system, the presence of strong SOC may also lead to the formation of flat band effect near the Fermi energy, showing the semiconducting ground state. Thus, both the resistivity and specific heat exhibit non-Fermi-liquid behavior at low temperatures in their normal states. Moreover, neither signature of semiconducting behaviors nor divergence of the specific heat coefficient has been observed for the pristine and slightly doped Sc$_2$Ir$_{3.9}$Si$_{0.1}$ sample at low temperatures in their normal states. However, it is found that the novel normal-state behavior also exists for Sc$_2$Ir$_{3.3}$Si$_{0.7}$, but not as pronounced as that for Sc$_2$Ir$_{3.5}$Si$_{0.5}$. The increase of the specific heat coefficient as $(\gamma_{\rm 0.45K}-\gamma_{\rm min})/\gamma_{\rm min}$ and the increase of the resistance as $(R_{\rm 2K}-R_{\rm min})/R_{\rm min}$ are shown in Fig.~\ref{fig1}(d). These are also in good agreement with the evolutions of $\gamma_n$ and $T_{\rm c}$ for the samples. Therefore, we can conclude that the strange non-Fermi-liquid behavior is strongly related to the evolutions of $T_{\rm c}$ and the effective DOS. We believe they should have the same origin.

Additionally, strong SCFs were observed for the samples around the optimal doping point. As can be seen in Fig.~\ref{fig2}(d), the superconducting transition of resistivity at the onset point is very round, and we define the critical temperature $T_{\rm {c}}^{*}\approx 5.7$ K at which the resistivity starts to drop. The enhanced conductivity extends up to $T_{\rm {c}}^{*}$ ($\sim$ 2$T_{\rm c}$) for Sc$_2$Ir$_{3.5}$Si$_{0.5}$, showing pronounced SCFs, being similar to previous reports on other unconventional superconductors \cite{56,57}. More strikingly, for Sc$_2$Ir$_{3.5}$Si$_{0.5}$, if the resistivity at 6 T is considered as the normal-state resistivity, the resistivity at 0 T starts to deviate from the ``normal state" at 18 K. This deviation point coincides with the temperature of the minimum resistivity. We may regard this point as the pseudogap temperature, which probably corresponds to the preformed Cooper pairs but without long-range phase coherence. This needs however more experimental verifications. Despite the large temperature region for the SCFs, the specific heat jump and the demagnetization transition near $T_{\rm c}$ are quite sharp with seemingly no trace of SCFs, this may be interpreted as the resolution limit on the tiny entropy contribution by the partial Cooper pairs in the SCF region. In the phase diagram of Fig.~\ref{fig1}(c), we also list the $T_{\rm {c}}^{*}$ from their resistivity measurements. Notably, the doping dependence of $\gamma_n$ in Fig.~\ref{fig1}(d) coincides with the doping dependence of $T_{\rm {c}}^{*}$ in Fig.~\ref{fig1}(c). Thus the enhancement of effective DOS ($\propto \gamma_n$) is closely related to the occurrence of the SCFs.

Usually, the normal-state specific heat at low temperatures obeys the Debye model with $C/T=\gamma + \beta T^2$ for a Fermi liquid ground state. However, we find that the specific heat coefficient deviates from the linear relationship with $T^2$ below 6 K, and the fitting to the normal-state specific heat of Sc$_2$Ir$_{3.5}$Si$_{0.5}$ by the Debye model yields a negative intercept of specific heat coefficient in the zero-temperature limit, as shown by the purple line in Fig.~\ref{fig3}(a). The Sommerfeld constant $\gamma_n$, which implies the effective DOS near the Fermi energy, must be positive. Considering the correlation effect mentioned above, in the fitting process, we introduce an additional contribution $-AT^{n}$ln$(T-\delta)$ which arises from the enhanced electron-electron interactions \cite{62}. The Debye model together with strong correlations shown as the blue line can fit the experimental data very well down to 0.5 K. The fitting parameter $n$ $\sim$ 2.2 $\textless$ 3 again confirms the strong correlations in this system. It is the direct experimental evidence for strong correlation effects from specific heat data \cite{63}.

We use the experimental data of specific heat at 6 T as the normal-state value and subtract it from the data at 0 T to obtain the electronic specific heat coefficient, the result is shown in Fig.~\ref{fig3}(b). Then we use the BCS formula with three different gap structures to fit it \cite{44}. Down to 0.5 K, the electronic specific heat coefficient decreases continuously without any trend of flattening. The fully gapped isotropic $s$-wave model plotted as the red line can be safely ruled out. The difference between extended $s$-wave and $d$-wave lies in the presence of energy gap nodes. It is not yet possible to directly determine the trend of electronic specific heat at very low temperatures. Thus, we cannot distinguish between the fits of the two gap structures indicated by the blue and green lines. However, we can expect that the normal state specific coefficient below 0.5 K should follow the increase trend with the logarithmic relation, as shown by the inset of Fig.~\ref{fig2}(b). In this case, the electronic specific heat coefficient related to superconductivity should continue to decrease without flattening at even lower temperatures. Thus, it indicates a strong possibility of $d$-wave pairing with gap nodes. The field-induced specific heat coefficients also support the possibility of $d$-wave pairing \cite{44}, but more direct evidence for its pairing symmetry is still lacking since we cannot accurately estimate the electronic specific heat coefficients in the zero-temperature limit.

\begin{figure}
  \includegraphics[width=8.8 cm]{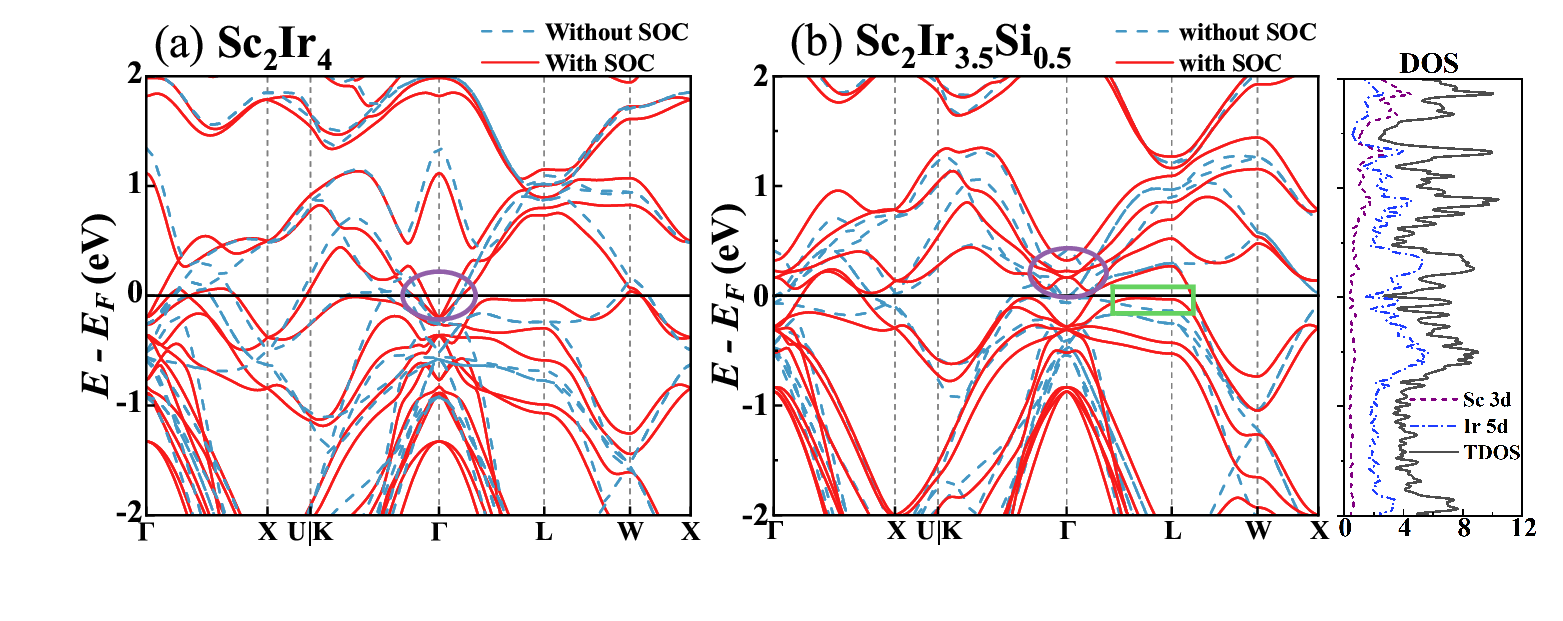}
\caption{Calculated electronic band structures around the Fermi energy ($E_{\rm F}$). Band structure without and with SOC for (a) Sc$_2$Ir$_4$ and for (b) Sc$_2$Ir$_{3.5}$Si$_{0.5}$. The green frame in (b) highlights the flat band between $\Gamma$ and L point. The panel on the right-hand side of (b) shows the partial and total DOS of Sc$_2$Ir$_{3.5}$Si$_{0.5}$ with SOC.} \label{fig4}
\end{figure}

Next we try to get some signatures of flat band effect from theoretical calculations of the electronic band structures, the results are shown in Fig.~\ref{fig4}. One can see that the dominant contribution to the DOS near Fermi energy $E_{\rm F}$ comes from the $5d$ electrons of Ir which play an important role in superconductivity \cite{44}. When the strong SOC of Ir is taken into account \cite{34,35}, large band splitting occurs, resulting in intriguing features. Initially, for the pristine sample, the two electron pockets around $\Gamma$ provide the main DOS, although it contains also a flat band between $\Gamma$-L after considering the SOC effect. However, after a small amount of Si doping, these two electron bands around $\Gamma$ are shifted upwards, leading to the decreased contribution of DOS at the Fermi energy. Meanwhile, a flat-like band still appears along the $\Gamma$-L direction near the Fermi level \cite{44}. It is this flat band that brings about large DOS with the enhanced effective mass of electrons. The combination of above mentioned two features, namely, less contribution of DOS from the Fermi pockets near $\Gamma$ and the enhancement of flat band effect after Si doping, can qualitatively explain the nonmonotonic doping dependence of $\gamma_n$ obtained from the specific heat measurements. Moreover, the estimated $\gamma_n$ is indeed consistent with the doping-dependent $T_{\rm c}$. Thus, the nonmonotonic doping dependence of DOS is responsible for the doping-dependent $T_{\rm c}$ which exhibits a second superconducting dome. In addition, correlations between electrons will gradually dominate due to the presence of the flat band, which also explains the deviation of the normal-state specific heat from the Debye model. To summarize, the shrinkage of the Fermi surface near $\Gamma$ and the presence of a flat band may cooperatively lead to enhanced correlation effect, resulting in the non-Fermi-liquid behavior in the normal state in the region of second superconducting dome.

In conclusion, we have successfully synthesized a superconducting system Sc$_2$Ir$_{4-x}$Si$_x$ with a phase diagram containing two superconducting domes. Many characteristics show that the superconductivity in the second dome is unconventional. This can be corroborated by the observations that (1) the normal states exhibit non-Fermi-liquid behaviors characterized by the logarithmic divergence of specific heat coefficient and the increased resistivity at low temperatures; (2) for Sc$_2$Ir$_{3.5}$Si$_{0.5}$ with the optimal $T_{\rm c}$, the superconducting-state specific heat shows the possible features of $d$-wave pairing; (3) strong superconducting fluctuations occur far above $T_{\rm c}$ at the optimal doping. Our theoretical calculations indicate that a flat band induced by the strong splitting after SOC can give a natural understanding of the tunable correlation effect, which can enhance $T_{\rm c}$ and induce strong superconducting fluctuations. All of the above findings provide new insights into the relationship between strong correlation effect and unconventional superconductivity in this interesting Iridium-based kagome system.

This work was supported by the National Key R$\and$D Program of China (No. 2022YFA1403201), National Natural Science Foundation of China (Nos. 12061131001, 52072170, 12204231, and 11927809), Strategic Priority Research Program (B) of Chinese Academy of Sciences (No. XDB25000000), and the Fundamental Research Funds for the Central Universities (Grant No. 020414380185).

 $\star$ These authors contributed equally to this work.

 Corresponding authors:

 $^\dagger$zhuxiyu@nju.edu.cn

 $^\ast$hhwen@nju.edu.cn


\begin{thebibliography}{00}
\bibitem{1}P. W. Anderson, Mater. Res. Bull. \textbf{8}, 153 (1973).
\bibitem{2}A. P. Ramirez, Annu. Rev. Mater. Sci. \textbf{24}, 453 (1994).
\bibitem{3}L. Balents, Nature (London) \textbf{464}, 199 (2010).
\bibitem{4}T.-H. Han, J S. Helton, S. Chu, D. G. Nocera, J. A. Rodriguez-Rivera, C. Broholm, and Y. S. Lee, Nature (London) \textbf{492}, 406 (2012).
\bibitem{5}S. V. Isakov, S. Wessel, R. G. Melko, K. Sengupta, and Y. B. Kim, Phys. Rev. Lett. \textbf{97}, 147202 (2006).
\bibitem{6}G.-W. Chern, P. Mellado, and O. Tchernyshyov, Phys. Rev. Lett. \textbf{106}, 207202 (2011).
\bibitem{7}M. Li, Q. Wang, G. Wang, Z. Yuan, W. Song, R. Lou, Z. Lu, Y. Huang, Z. Liu, H. Lei, Z. Yin, and S. Wang, Nat. Commun. \textbf{12}, 3129 (2021).
\bibitem{8}S. Cho, H. Ma, W. Xia, Y. Yang, Z. Liu, Z. Huang, Z. Jiang, X. Lu, J. Liu, Z. Liu, J. Li, J. Wang, Y. Liu, J. Jia, Y. Guo, J. Liu, and D. Shen, Phys. Rev. Lett. \textbf{127}, 236401 (2021).
\bibitem{9}S. Peng, Y. Han, G. Pokharel, J. Shen, Z. Li, M. Hashimoto, D. Lu, B. R. Ortiz, Y. Luo, H. Li, M. Guo, B. Wang, S. Cui, Z. Sun, Z. Qiao, S. D. Wilson, and J. He, Phys. Rev. Lett. \textbf{127}, 266401 (2021).
\bibitem{10}M. Kang, S. Fang, J.-K. Kim, B. R. Ortiz, S. H. Ryu, J. Kim, J. Yoo, G. Sangiovanni, D. D. Sante, B.-G. Park, C. Jozwiak, A. Bostwick, E. Rotenberg, E. Kaxiras, S. D. Wilson, J.-H. Park, and R. Comin, Nat. Phys. \textbf{18}, 301 (2022).
\bibitem{11}F. Laves and K. L\"{o}hberg, Nachr. Ges. Wiss. Goettingen \textbf{1}, 59 (1934).
\bibitem{12}F. Laves and H. Witte, Metallwirtsch \textbf{14}, 645 (1935).
\bibitem{13}S. B. Roy, Philos. Mag. B \textbf{65} 1435 (1992).
\bibitem{14}L. M. Schoop, L. S. Xie, R. Chen, Q. D. Gibson, S. H. Lapidus, I. Kimchi, M. Hirschberger, N. Haldolaarachchige, M. N. Ali, C. A. Belvin, T. Liang, J. B. Neaton, N. P. Ong, A. Vishwanath, and R. J. Cava, Phys. Rev. B \textbf{91}, 214517 (2015).
\bibitem{18}N. Haldolaarachchige, Q. Gibson, L. M. Schoop, H. Luo, and R. J. Cava, J. Phys.: Condens. Matter \textbf{27}, 185701 (2015).
\bibitem{19}R. Horie, K. Horigane, S. Nishiyama, M. Akimitsu, K. Kobayashi, S. Onari, T. Kambe, Y. Kubozono, and J. Akimitsu, J. Phys.: Condens. Matter \textbf{32}, 175703 (2020).
\bibitem{20}T. Koshinuma, H. Ninomiya, I. Hase, H. Fujihisa, Y. Gotoh, K. Kawashima, S. Ishida, Y. Yoshida, H. Eisaki, T. Nishio and A. Iyo, Intermetallics \textbf{148} 107643 (2022).
\bibitem{21}G. Xiao, S. Wu, B. Li, B. Liu, J. Wu, Y. Cui, Q. Zhu, G. Cao and Z. Ren, Intermetallics \textbf{128} 106993 (2021).
\bibitem{22}Q. S. Yang, B. B. Ruan, M. H. Zhou, Y. D. Gu, M. W. Ma, G. F. Chen and Z. A. Ren, Chin. Phys. B \textbf{32} 017402 (2023).
\bibitem{23}X. Yang, H. Li, T. He, T. Taguchi, Y. Wang, H. Goto, R. Eguchi, R. Horie, K. Horigane, K. Kobayashi, J. Akimitsu, H. Ishii, Y. F. Liao, H. Yamaoka and Y. Kubozono, J. Phys.: Condens. Matter \textbf{32} 025704 (2019).
\bibitem{24}S. Gutowska, K. G\'{o}rnicka, P. W\'{o}jcik, T. Klimczuk, and B. Wiendlocha, Phys. Rev. B \textbf{104}, 054505 (2021).
\bibitem{25}Y. Zhang, X. M. Tao and M. Q. Tan, Chin. Phys. B \textbf{26} 047401 (2017).
\bibitem{33}T. H. Geballe, B. T. Matthias, V. B. Compton, E. Corenzwit, G. W. Hull, and L. D. Longinotti, Phys. Rev. \textbf{137}, A119 (1965).
\bibitem{34}H. Y. Uzunok, Sakarya University Journal of Science, \textbf{24}, 406 (2020).
\bibitem{35}U. K. Chowdhury and T. C. Saha, Phys. Solid State \textbf{61}, 530 (2019).
\bibitem{definefb}N. Regnault, Y. Xu, M.-R. Li, D.-S. Ma, M. Jovanovic, A. Yazdani, S. S. P. Parkin, C. Felser, L. M. Schoop, N. P. Ong, R. J. Cava, L. Elcoro, Z.-D. Song, and B. A. Bernevig, Nature (London) \textbf{603}, 824 (2022).
\bibitem{flatband}L. Balents, C. R. Dean, D. K. Efetov, and A. F. Young, Nat. Phys. \textbf{16}, 725 (2020).
\bibitem{CsVSbflat}Y. Hu, S. M. L. Teicher, B. R. Ortiz, Y. Luo, S. Peng, L. Huai, J. Ma, N. C. Plumb, S. D. Wilson, J. He, and M. Shi, Sci. Bull. \textbf{67}, 495 (2022).
\bibitem{CsTiBiflat}J. Yang, X. Yi, Z. Zhao, Y. Xie, T. Miao, H. Luo, H. Chen, B. Liang, W. Zhu, Y. Ye, J.-Y. You, B. Gu, S. Zhang, F. Zhang, F. Yang, Z. Wang, Q. Peng, H. Mao, G. Liu, Z. Xu, H. Chen, H. Yang, G. Su, H. Gao, L. Zhao, and X. J. Zhou, Nat. Commun. \textbf{14}, 4089 (2023).
\bibitem{SOCfb}C. Weeks and M. Franz, Phys. Rev. B \textbf{85}, 041104(R) (2012).
\bibitem{64}M. Imada and M. Kohno, Phys. Rev. Lett. \textbf{84}, 143 (2000).
\bibitem{66}V. J. Kauppila, F. Aikebaier, and T. T. Heikkil\"{a}, Phys. Rev. B \textbf{93}, 214505 (2016).
\bibitem{LaRuSi}C. Mielke, Y. Qin, J.-X. Yin, H. Nakamura, D. Das, K. Guo, R. Khasanov, J. Chang, Z. Q. Wang, S. Jia, S. Nakatsuji, A. Amato, H. Luetkens, G. Xu, M. Z. Hasan, and Z. Guguchia, Phys. Rev. Mater. \textbf{5}, 034803 (2021).
\bibitem{65}Y. He, S.-D. Chen, Z.-X. Li, D. Zhao, D. Song, Y. Yoshida, H. Eisaki, T. Wu, X.-H. Chen, D.-H. Lu, C. Meingast, T. P. Devereaux, R. J. Birgeneau, M. Hashimoto, D.-H. Lee, and Z.-X. Shen, Phys. Rev. X \textbf{11}, 031068 (2021).
\bibitem{Mag}Y. Cao, V. Fatemi, A. Demir, S. Fang, S. L. Tomarken, J. Y. Luo, J. D. Sanchez-Yamagishi, K. Watanabe, T. Taniguchi, E. Kaxiras,  R. C. Ashoori, and P. Jarillo-Herrero, Nature (London) \textbf{556}, 80 (2018).
\bibitem{NiIn}L. Ye, S. Fang, M. G. Kang, J. Kaufmann, Y. Lee, J. Denlinger, C. Jozwiak, A. Bostwick, E. Rotenberg, E. Kaxiras, D. C. Bell, O. Janson, R. Comin, and J. G. Checkelsky, arXiv:2106.10824.
\bibitem{44}See Supplemental Material
\bibitem{15}L. Siggelkow, V. Hlukhyy, and T. F. F\"{a}ssler, Z. Anorg. Allg. Chem. \textbf{643} 1424 (2017).
\bibitem{16}K. Cenzual, B. Chabot and E. Parth\'{e}, J. Solid State Chem. \textbf{70} 229 (1987).
\bibitem{17}V. Keimes and A. Mewis, Z. Naturforsch. B \textbf{47} 1351 (1992).
\bibitem{45}J. W. Garland, K. H. Bennemann, and F. M. Mueller, Phys. Rev. Lett. \textbf{21}, 1315 (1968).
\bibitem{HEA}L. Sun and R. J. Cava, Phys. Rev. Mater. \textbf{3}, 090301 (2019).
\bibitem{46}H. Q. Yuan, F. M. Grosche, M. Deppe, C. Geibel, G. Sparn, and F. Steglich, Science \textbf{302}, 2104 (2003).
\bibitem{47}K. Y. Chen, N. N. Wang, Q. W. Yin, Y. H. Gu, K. Jiang, Z. J. Tu, C. S. Gong, Y. Uwatoko, J. P. Sun, H. C. Lei, J. P. Hu, and J.-G. Cheng, Phys. Rev. Lett. \textbf{126}, 247001 (2021).
\bibitem{48}S. Iimura, S. Matsuishi, H. Sato, T. Hanna, Y. Muraba, S. W. Kim, J. E. Kim, M. Takata, and H. Hosono, Nat. Commun. \textbf{3}, 943 (2012).
\bibitem{49}P. Reiss, D. Graf, A. A. Haghighirad, W. Knafo, L. Drigo, M. Bristow, A. J. Schofield, and A. I. Coldea, Nat. Phys. \textbf{16}, 89 (2020).
\bibitem{cuprates}M. Kofu, S.-H. Lee, M. Fujita, H.-J. Kang, H. Eisaki, and K. Yamada, Phys. Rev. Lett. \textbf{102}, 047001 (2009).
\bibitem{50}K. A. Moler, D. L. Sisson, J. S. Urbach, M. R. Beasley, A. Kapitulnik, D. J. Baar, R. Liang, and W. N. Hardy, Phys. Rev. B \textbf{55}, 3954 (1997).
\bibitem{51}J. E. Gordon, R. A. Fisher, Y. X. Jia, N. E. Phillips, S. F. Reklis, D. A. Wright, and A. Zettl, Phys. Rev. B \textbf{59}, 127 (1999).
\bibitem{52}J. P. Emerson, R. A. Fisher, N. E. Phillips, D. A. Wright and E. M. McCarron III, Phys. Rev. B \textbf{49}, 9256 (1994).
\bibitem{53}H. v. L\"{o}hneysen, A. Rosch, M. Vojta, and P. W\"{o}lfle, Rev. Mod. Phys. \textbf{79}, 1015 (2007).
\bibitem{54}T. Shibauchi, A. Carrington, and Y. Matsuda, Annu. Rev. Condens. Matter Phys. \textbf{5}, 113 (2014).
\bibitem{55}B. Michon, C. Girod, S. Badoux, J. Ka\v{c}mar\v{c}\'{i}k, Q. Ma, M. Dragomir, H. A. Dabkowska, B. D. Gaulin, J.-S. Zhou, S. Pyon, T. Takayama, H. Takagi, S. Verret, N. Doiron-Leyraud, C. Marcenat, L. Taillefer, and T. Klein, Nature (London) \textbf{567}, 218 (2019).
\bibitem{CeRhGe}B. Shen, Y. Zhang, Y. Komijani, M. Nicklas, R. Borth, A. Wang, Y. Chen, Z. Nie, R. Li, X. Lu, H. Lee, M. Smidman,
F. Steglich, P. Coleman, and H. Yuan, Nature (London) \textbf{579}, 51 (2020).
\bibitem{HF}G. R. Stewart, Rev. Mod. Phys. \textbf{56}, 755 (1984).
\bibitem{58}J. Kondo, Progress of Theoretical Physics \textbf{32} 37 (1964)
\bibitem{59}X. Ding, J. Xing, G. Li, L. Balicas, K. Gofryk, and H. H. Wen, Phys. Rev. B \textbf{103}, 125115 (2021).
\bibitem{60}P. W. Anderson, Phys. Rev. \textbf{109}, 1492 (1958).
\bibitem{61}D. J. Thouless, Phys. Rep. \textbf{13}, 93 (1974).
\bibitem{38}G. S. Boebinger, Y. Ando, A. Passner, T. Kimura, M. Okuya, J. Shimoyama, K. Kishio, K. Tamasaku, N. Ichikawa, and S. Uchida, Phys. Rev. Lett. \textbf{77}, 5417 (1996).
\bibitem{39}S. Komiya, H.-D. Chen, S.-C. Zhang, and Y. Ando, Phys. Rev. Lett. \textbf{94}, 207004 (2005).
\bibitem{29}M. Xie and A. H. MacDonald, Phys. Rev. Lett. \textbf{124}, 097601 (2020).
\bibitem{30}G. Sethi, Y. Zhou, L. Zhu, L. Yang, and F. Liu, Phys. Rev. Lett. \textbf{126}, 196403 (2023).
\bibitem{56}H. Yang, G. Chen, X. Zhu, J. Xing, and H.-H. Wen, Phys. Rev. B \textbf{96}, 064501 (2017).
\bibitem{57}F. Rullier-Albenque, H. Alloul, and G. Rikken, Phys. Rev. B \textbf{84}, 014522 (2011).
\bibitem{62}H. v. L\"{o}hneysen, A. Rosch, M. Vojta, and P. W\"{o}lfle, Rev. Mod. Phys. \textbf{79}, 1015 (2007).
\bibitem{63}S. Li, B. Zeng, X. Wan, J. Tao, F. Han, H. Yang, Z. Wang, and H.-H. Wen, Phys. Rev. B \textbf{84}, 214527 (2011).
\end{thebibliography}
\end{document}

\title{Unconventional superconductivity in Sc$_2$Ir$_{4-x}$Si$_x$ by spin-orbit coupling driven flat band}
\author{Zhengyan Zhu$^{\star}$, Yuxiang Wu$^{\star}$, Shengtai Fan, Yiliang Fan, Yiwen Li, Yongze Ye, Xiyu Zhu$^{\dag}$, Haijun Zhang, Hai-Hu Wen$^{\ast}$}

\affiliation{Center for Superconducting Physics and Materials, National Laboratory of Solid State Microstructures and Department of Physics,  Collaborative Innovation Center for Advanced Microstructures, Nanjing University, Nanjing 210093, China}
\date{\today}
\maketitle

\section{Sample preparations}
The samples of Sc$_2$Ir$_{4-x}$Si$_x$ ($x=0 \sim 0.7$) were synthesized by the arc-melting method. The raw materials scandium (99.9$\%$, 22 mesh, Alfa Aesar), iridium (99.5$\%$, 325 mesh, Alfa Aesar), and silicon (99.99$\%$, 100 mesh, Alfa Aesar) were weighed in stoichiometric proportions, grounded and then pressed into a pellet in a glove box (water moisture and the oxygen compositions are below 0.1 PPM). To obtain a uniform sample, every pellet was melted at least three times and flipped over at intermediate intervals. The resulting air-stable pellets were wrapped with Ta sheets, sealed in an evacuated quartz tube under a high vacuum, and annealed at 800 $^\circ$C for 4 days.
\section{Experimental details}
The crystal structure was evaluated via XRD (Bruker, D8 Advance diffractometer) with Cu K$\alpha$ radiation ($\lambda=1.5418$ $\AA$). The intensity data were obtained over the $2\theta$ range of 10$^\circ$ to 100$^\circ$, with a step width of 0.01$^\circ$. The Rietveld refinements were conducted with TOPAS 4.2 software \cite{1}. The dc magnetization measurements were performed with a SQUID-VSM-7 T down to 1.8 K (Quantum Design). The electrical resistivity measurements were measured by employing a standard four-probe method with a physical property measurement system (PPMS 16 T, Quantum Design). The specific heat was measured by the thermal-relaxation method with an additional $^3$He insert using the PPMS-16 T. To calculate the electronic band structure with and without the spin-orbit coupling, we performed the first-principles calculations using Vienna Ab initio Simulation Package (VASP) \cite{2,3}. The generalized gradient approximation (GGA) in the Perdew-Burke Ernzerhof form \cite{4} was used in the self-consistently calculation of the electronic structure. Here, we set a $9\times9\times9$ k mesh with a cutoff energy of 370 eV, and the virtual-crystal approximation (VCA) was used to take into account the effect of partial substitution of Si for Ir.
\section{Sample characterization}
Figure~\ref{figs0}(a) shows the crystal structure of Sc$_2$Ir$_{4-x}$Si$_x$, which belongs to the $C15$ type laves phase with a space group of $Fd-3m$. For the pristine, every four Ir atoms form a tetrahedra which is connected through sharing vertices. In one unit cell, four small tetrahedra of Ir occupy the vertex positions of a larger tetrahedra. By the arc melting method, we synthesized a series of compounds Sc$_2$Ir$_{4-x}$Si$_x$ ($x=0\sim0.7$). Here, Si atoms randomly and disorderly replace the Ir atoms. The Fig.~\ref{figs0}(b) presents powder X-ray diffraction (XRD) patterns and their crystallinity index for Sc$_2$Ir$_{4-x}$Si$_x$ ($x=0\sim0.7$). As we can see, for the doping level less than or equal to 0.5, the peaks are very sharp and samples have good crystallinity. When the doping level reaches 0.7, more impurity of ScIr$_3$ appears in the sample, and the XRD peaks become broader, which means that the crystallinity of the sample deteriorates. Therefore, the doping level of Si in ScIr$_2$ is limited. The Fig.~\ref{figs0}(c) shows the powder XRD pattern and corresponding Rietveld refinements of the polycrystalline sample of Sc$_2$Ir$_{3.5}$Si$_{0.5}$. The sample is quite pure and can be well-fitted by the ScIr$_2$ structure with only tiny non-superconducting impurities of ScIr$_3$. These confirm the high quality of the polycrystalline samples from pristine to $x = 0.5$. Meanwhile, as the doping level increases, the peaks of XRD shift to the right as shown in Fig.~\ref{figs0}(d), indicating a decrease in lattice constant. In Fig.~\ref{figs0}(e), we show the variation of lattice constants with doping levels. The lattice constant decreases with increasing the doping level. The decrease in lattice constant is due to the much smaller atomic radius of Si compared to Ir.
\renewcommand\thefigure{S\arabic{figure}}
\setcounter{figure}{0}
\begin{figure}[H]
  \centering
  \vspace{-1cm}
  \includegraphics[trim=10 0 0 10, width=15cm]{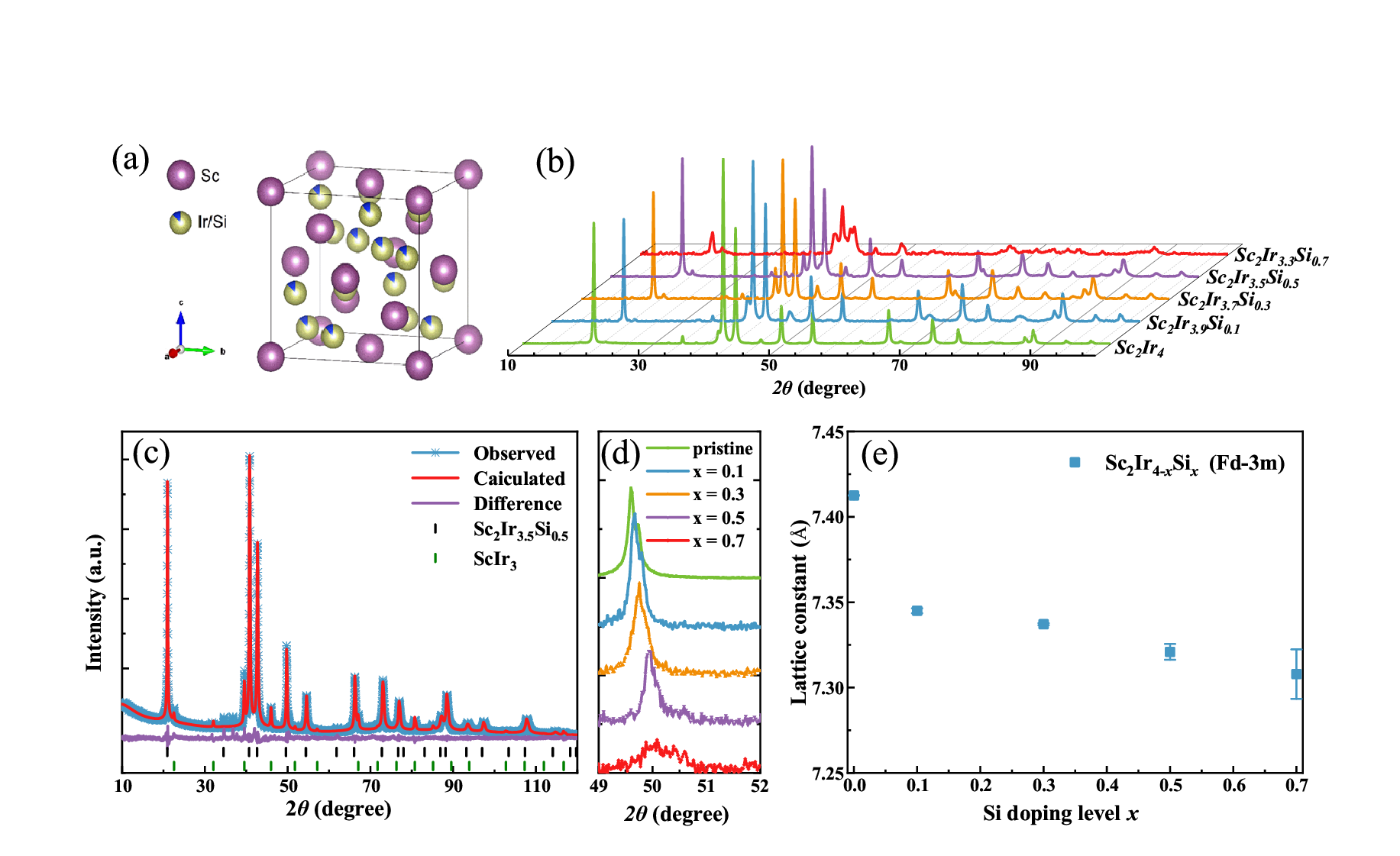}
  \caption{(a) The crystal structure of Sc$_2$Ir$_{4-x}$Si$_x$. (b) The 3D powder XRD patterns for series samples Sc$_2$Ir$_{4-x}$Si$_x$ from $x=0$ to $x=0.7$. The height means normalized intensity here. (c) The Powder XRD pattern (blue line) and Rietveld fitting curve (red line) of Sc$_2$Ir$_{3.5}$Si$_{0.5}$. (d) The shift of the peaks for the XRD patterns of series Sc$_2$Ir$_{4-x}$Si$_x$. (e) The extracted lattice parameters as a function of Si doping level $x$ for Sc$_2$Ir$_{4-x}$Si$_x$.} \label{figs0}
\end{figure}
\section{The upper critical field of the sample Sc$_2$Ir$_{3.5}$Si$_{0.5}$}
\begin{figure}[H]
  \centering
  \vspace{-0.5cm}
  \includegraphics[width=15cm]{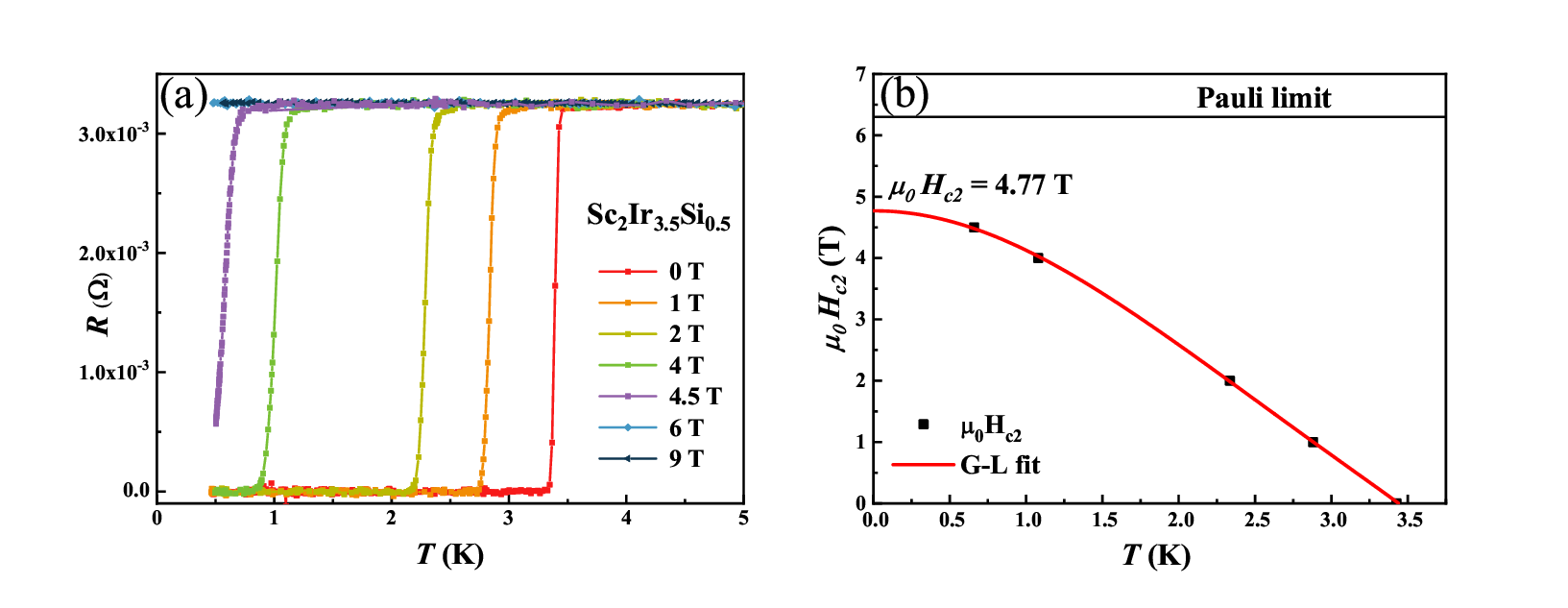}
\caption{(a) The temperature dependence of the resistance for Sc$_2$Ir$_{3.5}$Si$_{0.5}$ at various fields up to 9 T. (b) The temperature-dependent upper critical field of Sc$_2$Ir$_{3.5}$Si$_{0.5}$ and the fitting result by the Ginzburg-Landau (GL) theory.} \label{figs1}
\end{figure}

As shown in Fig.~\ref{figs1}(a), with the increasing magnetic field, the superconducting transition temperature decreases gradually. At 6 T and 9 T, there is no superconducting phenomenon down to 0.4 K. We collect the temperature dependence of the upper critical field with the criteria of 90$\%$ of the normal-state resistivity from the $R-T$ curves. The data shown in Fig.~\ref{figs1}(b) fit well by the Ginzburg-Landau theory $H_{\rm{c2}}(T)=H_{\rm c2}(0)(1-(T/T_{\rm c})^2)^a/(1+(T/T_{\rm c})^2)^b$ with the $\mu_{\rm 0}H_{\rm {c2}}(0)=$ 4.77 T lower than weak-coupling Pauli limit. In the series of samples, Sc$_2$Ir$_{3.5}$Si$_{0.5}$ has the highest $T_{\rm c}$, so the upper critical field of all other samples also cannot exceed 5 T. This strongly proves that all samples are in the normal state at 6 T and 9 T.
\section{The magnetization behavior of Sc$_2$Ir$_{3.5}$Si$_{0.5}$}
\begin{figure}[H]
  \centering
  \includegraphics[width=8cm]{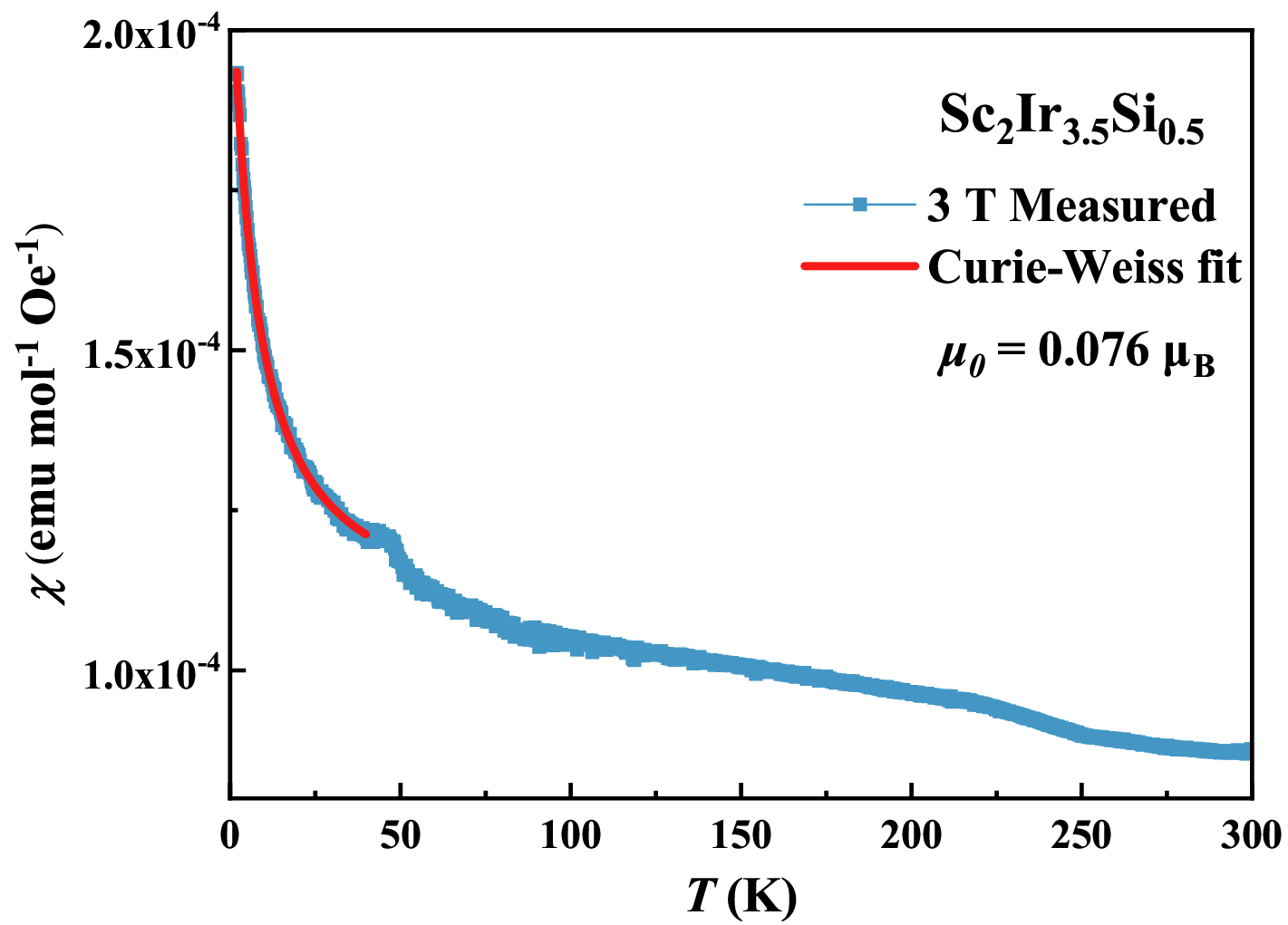}
\caption{The temperature dependence of magnetic susceptibility measured at 3 T and its Curie-Weiss fit.} \label{figs2}
\end{figure}

To investigate the magnetization behavior in the normal state, here shows the temperature-dependent magnetic susceptibility under a high field down to 2 K in Fig.~\ref{figs2}. The peak around 50 K results from the frozen oxygen. No clear long-range magnetic order is observed in the whole measured temperature range. In the low-temperature region, it exhibits paramagnetic Curie-Weiss behavior. The Curie Weiss law is used to fit the curve by the equation $\chi(T)=\chi_0+N \mu_0 \mu_J^2 / 3k_B (T+T_{\theta})$ from 2 to 40 K. We can derive the effective magnetic moment $\mu_0 = 0.076 \mu_{\rm B}$. Such a small magnetic moment rules out the possibility of localized magnetic impurities \cite{5}.
\section{The magnetic susceptibility and specific heat measurements for series samples}
\begin{figure}[H]
  \centering
    \vspace{-1cm}
  \includegraphics[width=15cm]{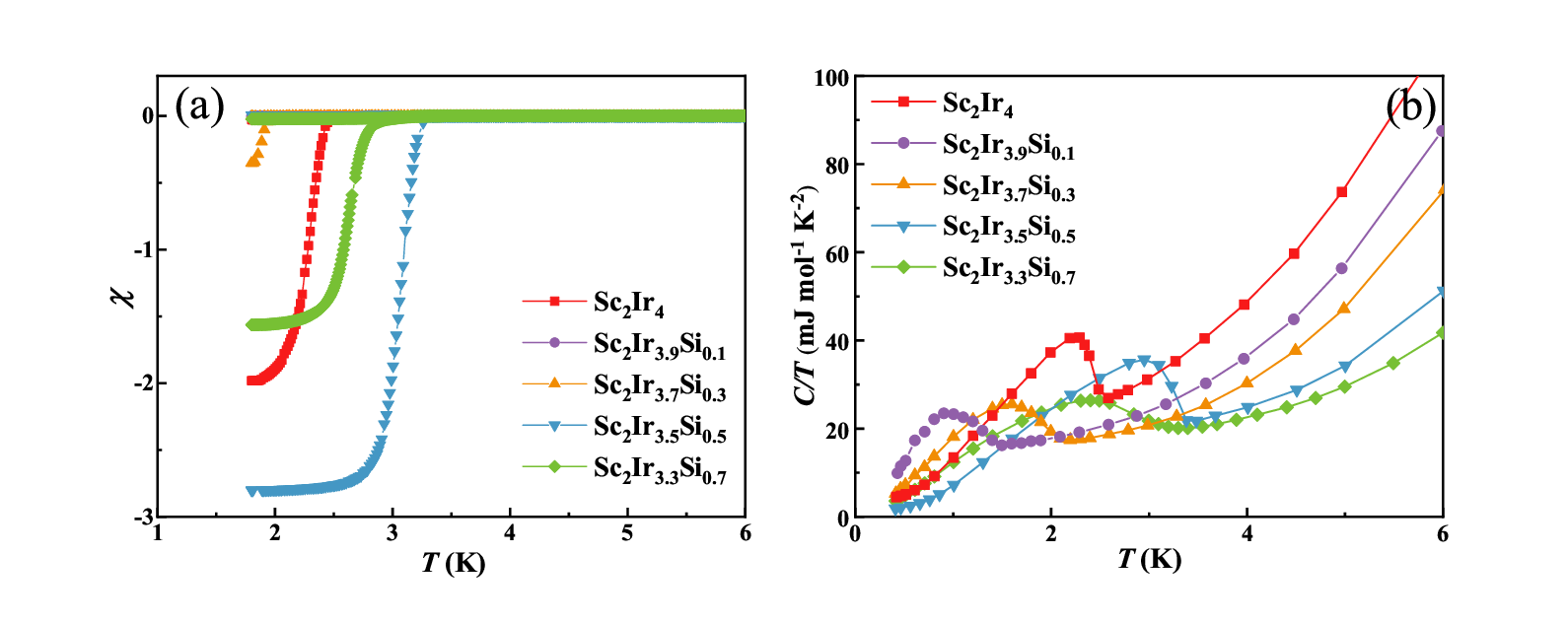}
\caption{ (a) The temperature-dependent magnetic susceptibility of Sc$_2$Ir$_{4-x}$Si$_x$ with an applied magnetic field of 10 Oe. (b) The temperature dependence of specific heat for series samples from 0.4 K to 6 K at 0 T.} \label{figs3}
\end{figure}

Figure~\ref{figs3}(a) shows the magnetic susceptibility and Figure~\ref{figs3}(b) shows the specific heat for corresponding samples at 0 T. For the doping levels less than and equal to 0.5, sharp superconducting transitions can be seen around $T_{\rm c}$. While for Sc$_2$Ir$_{3.3}$Si$_{0.7}$, the purity and crystallinity of the sample decrease. The diamagnetic transition and the specific heat jump both become broader. Considering the demagnetization factor, for most of the samples, their superconducting volumes are above 100$\%$. Only for the sample with $x=0.3$ shows an incomplete superconducting transition down to 1.8 K. Moreover, if extend the specific heat curves to zero temperature, we can speculate that their interceptions are small \cite{6}. The small residual specific heat coefficients imply that there are few impurities in the samples. All above are consistent with the XRD and electrical transport properties, once again proving the bulk superconductivity and high quality of the samples.
\section{The specific heat and resistance measurements for other doping levels}
\begin{figure}[H]
  \centering
  \includegraphics[width=14cm]{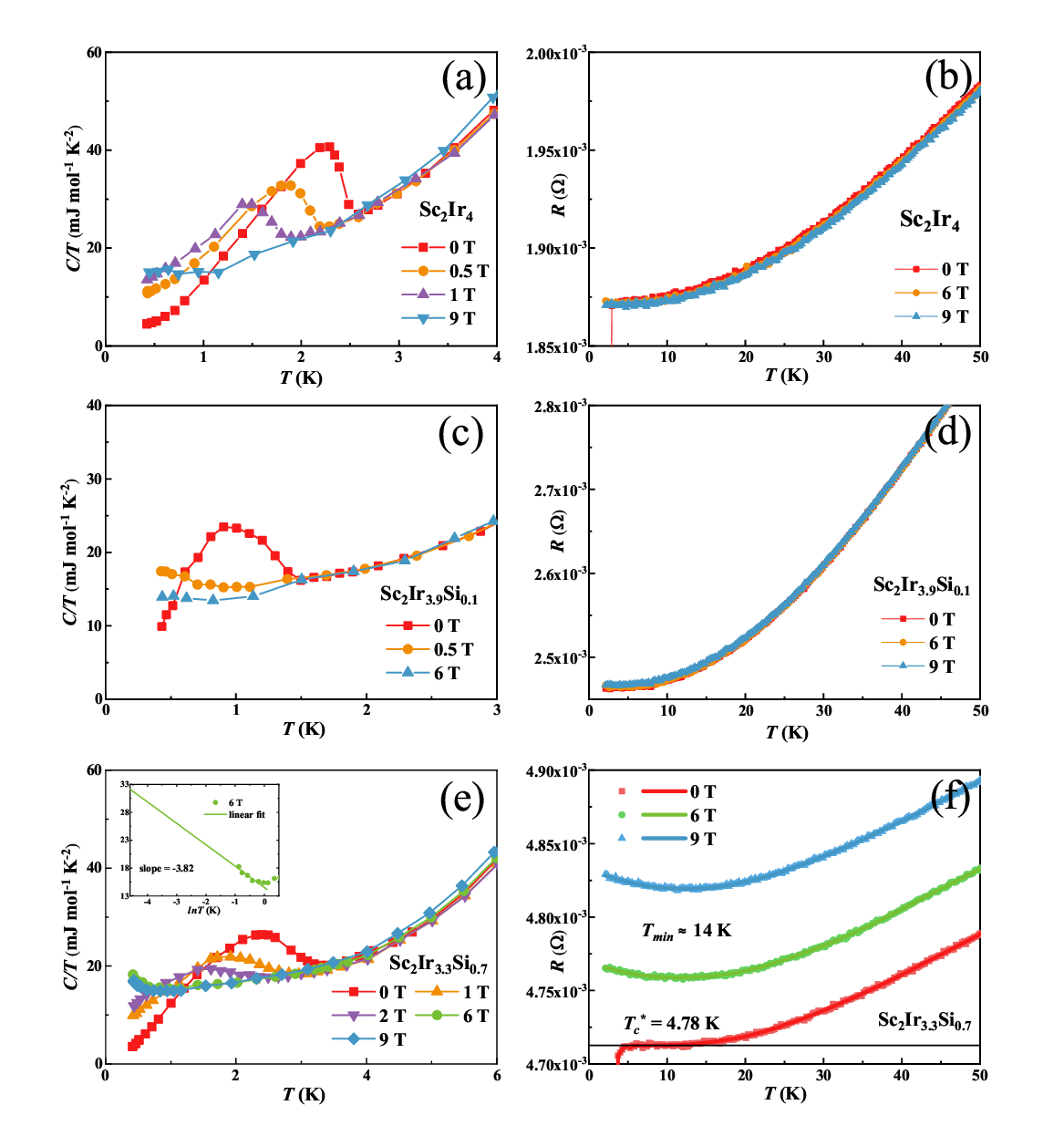}
\caption{The temperature dependence of specific heat (a)(c)(e) and resistance (b)(d)(f) under various magnetic fields for the pristine Sc$_2$Ir$_4$, Sc$_2$Ir$_{3.9}$Si$_{0.1}$ and Sc$_2$Ir$_{3.3}$Si$_{0.7}$ respectively.} \label{figs4}
\end{figure}

Figure~\ref{figs4}(a) and ~\ref{figs4}(c) present the specific heat under different magnetic fields for the pristine Sc$_2$Ir$_4$ and Sc$_2$Ir$_{3.9}$Si$_{0.1}$. Sharp specific heat jump can be seen at 0 T showing the high quality of the samples. In the normal state, specific heat continues to decrease down to 0.4 K showing no evidence of upturn or divergence with the temperature decreases. Fig.~\ref{figs4}(b) and ~\ref{figs4}(d) show the temperature dependence of the resistance for the pristine Sc$_2$Ir$_4$ and Sc$_2$Ir$_{3.9}$Si$_{0.1}$ at 0 T, 6 T, and 9 T, respectively. Both samples display metallic behavior above the critical temperature of superconductivity. And they behave like Landau Fermi liquid down to 2 K in their normal state. No signals of any little increase of resistance exist when coming to low temperatures.

Figure~\ref{figs4}(e) presents the specific heat under different magnetic fields for Sc$_2$Ir$_{3.3}$Si$_{0.7}$. Clear upturn of the specific heat coefficient can be seen at low temperatures in the normal state. Additionally, the resistance also increases below 14 K under high magnetic fields. The divergence of specific heat coefficient and the rise of the resistance both are lower than those for Sc$_2$Ir$_{3.5}$Si$_{0.5}$. As a result of the lower crystal crystallinity, the resistance under magnetic fields exhibits visible magnetoresistance. Considering that the sample with $x=0.7$ has reached the doping limit and the crystallinity of the sample is also low, the measurements of Sc$_2$Ir$_{3.3}$Si$_{0.7}$ are only for reference.
\section{The optimized fits according to different gap structures for series samples}
Here shows the electronic specific heat coefficient obtained by subtracting the normal state specific heat from the total specific heat data. And we plot them as $(C-C_n)/T$ versus $T$. We try the BCS formula with different gap structures to fit the electronic specific heat coefficient for all other samples. Here list the different gap structures adopted here: a single $s$-wave gap $\Delta(T,\theta) = \Delta_0 (T)$ as red lines, a single $d$-wave gap $\Delta(T,\theta) = \Delta_0 (T)cos2\theta$ as blue lines, and a single extended $s$-wave gap $\Delta(T,\theta) = \Delta_0(T)(1 + cos2\theta)$ as green lines and $\alpha$ represents the anisotropy of the energy gap.

\begin{figure}[H]
\centering
  \includegraphics[width=15cm]{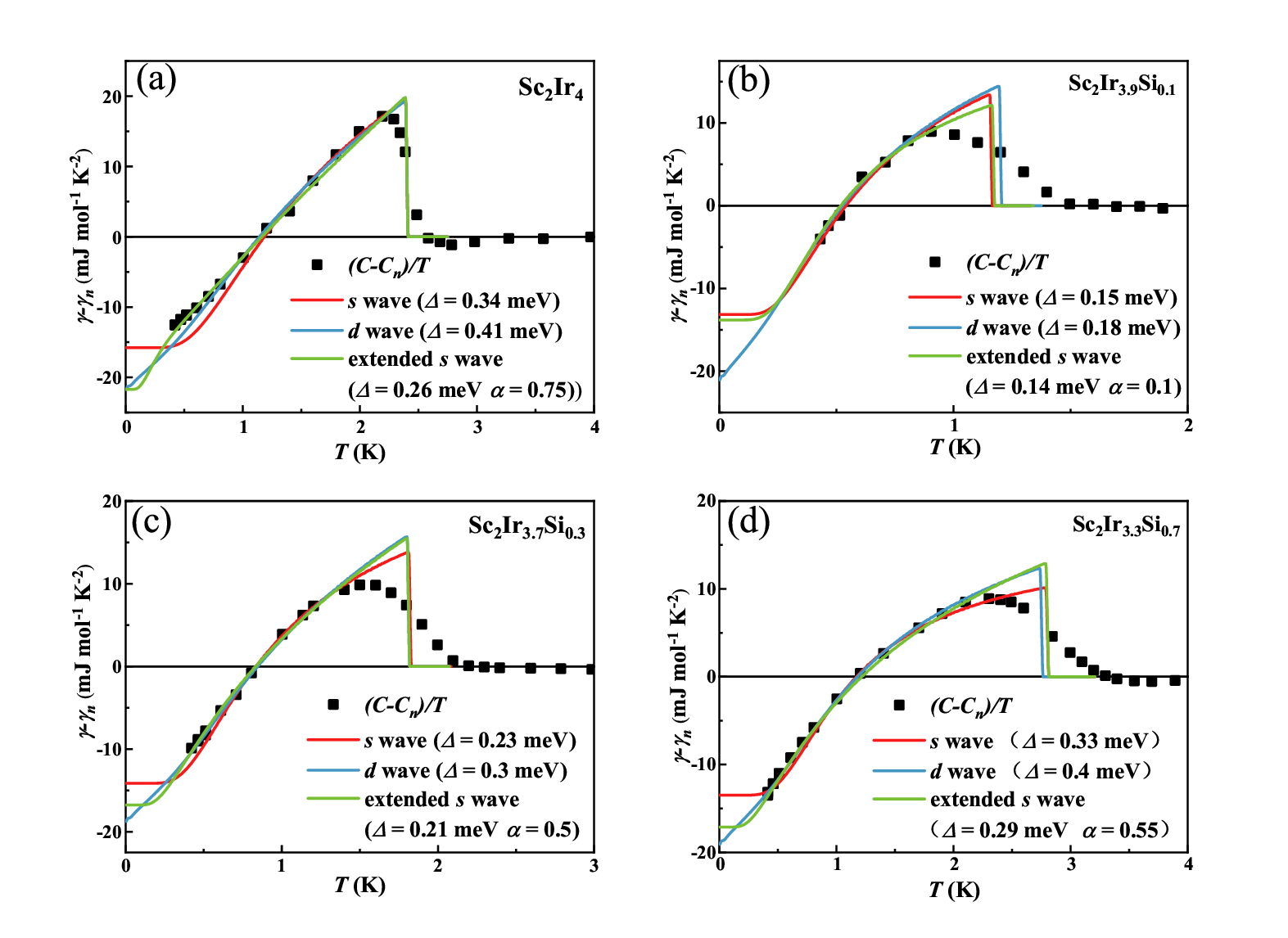}
\caption{ The temperature dependence of electronic specific heat coefficient at zero field (symbols) and the fitting curves of different gap structures (a) for the pristine sample, (b) for Sc$_2$Ir$_{3.9}$Si$_{0.1}$, (c) for Sc$_2$Ir$_{3.7}$Si$_{0.3}$ and (d) for Sc$_2$Ir$_{3.3}$Si$_{0.7}$.} \label{figs5}
\end{figure}

Considering a $d$-wave gap structure, it exists gap nodes. Around the gap nodes, quasiparticles can be easily excited. But for the $s$-wave case, due to the uniform energy gap, quasiparticles require more energy to excite. So, in this case, there may present a small plateau for its specific heat coefficient down to the low-temperature limit. And for the anisotropic $s$-wave gap, although fully gapped, there are energy gap minima. The specific heat coefficient can continue to decrease down to lower temperatures than the $s$-wave. Our specific heat data all decrease until 0.4 K, which may not match the $s$-wave energy gap. For the sample Sc$_2$Ir$_{3.7}$Si$_{0.3}$ in Fig.~\ref{figs5}(c) and Sc$_2$Ir$_{3.3}$Si$_{0.7}$ in Fig.~\ref{figs5}(d), the $d$-wave gap model and extended $s$-wave gap model both can fit the data well. The calculated $2\Delta/k_{\rm B} T_{\rm c}$ under the $d$-wave model is 3.87 for Sc$_2$Ir$_{3.7}$Si$_{0.3}$ and 3.41 for Sc$_2$Ir$_{3.3}$Si$_{0.7}$. With the extended $s$-wave gap, the calculated $2\Delta/k_{\rm B} T_{\rm c}$ is 2.69 for Sc$_2$Ir$_{3.7}$Si$_{0.3}$ and 2.4 for Sc$_2$Ir$_{3.3}$Si$_{0.7}$. For the pristine, the extended $s$-wave fit best with $2\Delta/k_{\rm B} T_{\rm c}$ about 2.51, and the others fittings both deviate at low temperatures as can be seen in Fig.~\ref{figs5}(a). And for Sc$_2$Ir$_{3.9}$Si$_{0.1}$, the $T_{\rm c}$ of the sample is too low to distinguish the three different gap structures shown in Fig.~\ref{figs5}(b). The calculated $2\Delta/k_{\rm B} T_{\rm c}$ all below the weak-coupling BCS theory 3.53 for $s$-wave and 4.28 for $d$-wave. Such weak electron-phonon coupling suggests that the existence of another superconducting pairing mechanism that goes beyond conventional superconductors.
\section{The magnetic-field induced specific heat coefficient at zero temperature limit for Sc$_2$Ir$_{3.5}$Si$_{0.5}$}
\begin{figure}[H]
\centering
  \includegraphics[width=12cm]{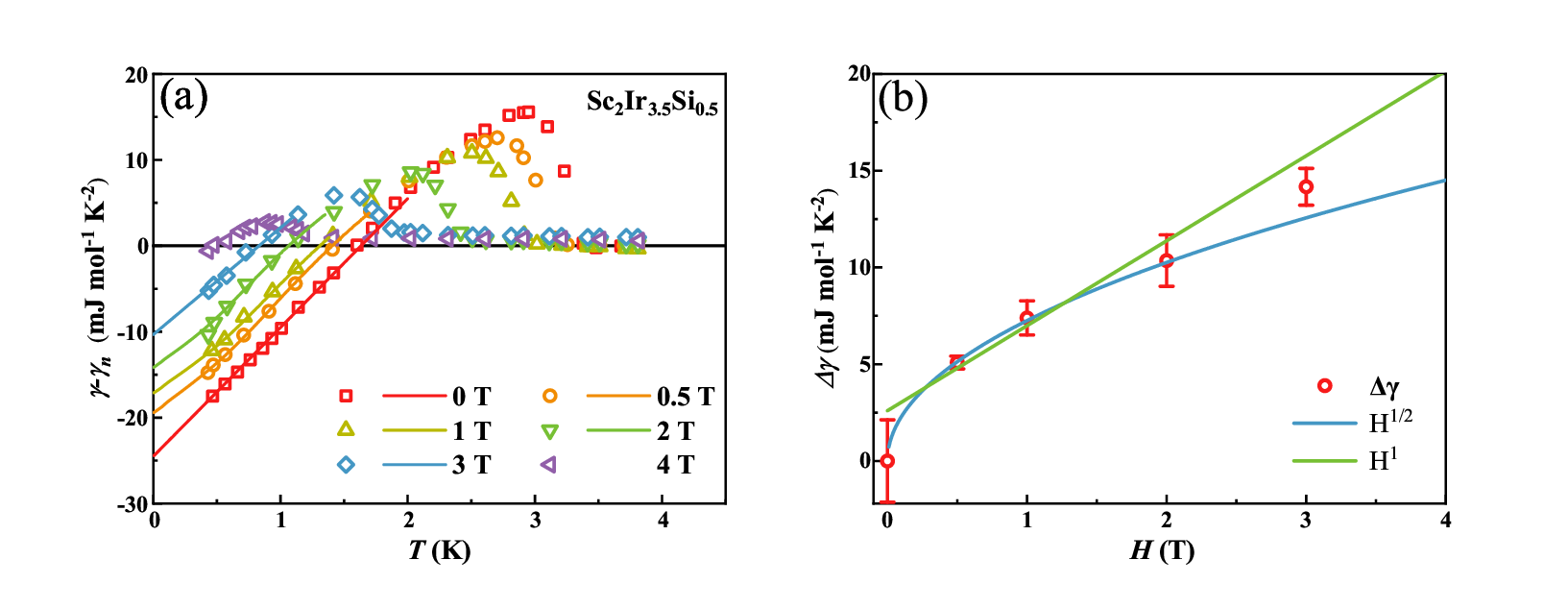}
\caption{The temperature dependence of electronic specific heat coefficient under different magnetic fields for Sc$_2$Ir$_{3.5}$Si$_{0.5}$. (b) The magnetic field-dependent specific heat coefficient and two kinds of fits by different power laws.} \label{figs6}
\end{figure}

We obtain the electronic specific heat coefficient under different magnetic fields for Sc$_2$Ir$_{3.5}$Si$_{0.5}$ by subtraction of the normal state specific heat data at 6 T, as shown by symbols in Fig.~\ref{figs6}(a). The lines are the linear extensions of the specific heat data down to zero temperature limit to acquire the electronic specific heat coefficient at 0 K. The magnetic field-dependent specific heat coefficients are in Fig.~\ref{figs6}(b) as symbols with error bars by subtracting the data at 0 T.

The field-induced specific heat coefficient is also related to the energy gap structure. For the $s$-wave model, the field-induced specific heat only arises from the increase of the number of magnetic flux, and the $\Delta \gamma$ is proportional to $H$. For the $d$-wave model with gap nodes, the field-induced specific heat is additionally influenced by the quasiparticles around the magnetic flux due to the Doppler effect, and the $\Delta \gamma$ is proportional to $H^{1/2}$. Here, we adopt the two kinds of power law to fit the magnetic field-dependent specific heat coefficient. At low magnetic fields, the field-induced specific heat slightly deviates from the linear fit with a negative curvature. Thus, the data fit better by the $H^{0.5}$ than the $H^{1}$, suggesting the possible gap nodes in the gap structure. The field-induced specific heat gives another proof about the gap structure and again supports the $d$-wave model for the gap structure. However, due to the inability to estimate the low-temperature specific heat trend, the electronic specific heat coefficient at 0 K obtained by extrapolation contains large errors, and therefore this conclusion requires strong justification from other measurements.
\section{The calculated band structures for Sc$_2$Ir$_{3.5}$Si$_{0.5}$ with SOC}
\begin{figure}[H]
 \centering
  \setlength{\abovecaptionskip}{-0.1cm}
  \setlength{\belowcaptionskip}{-0.5cm}
  \includegraphics[width=6cm]{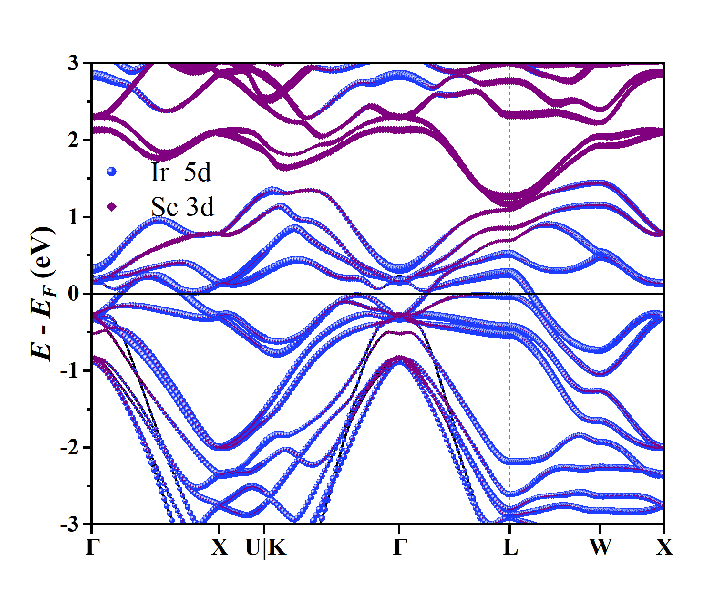}
\caption{The calculated band structures for Sc$_2$Ir$_{3.5}$Si$_{0.5}$ with SOC} \label{figs7}
\end{figure}

Figure~\ref{figs7} shows the calculated electronic band structure for Sc$_2$Ir$_{3.5}$Si$_{0.5}$ with strong spin-orbit coupling (SOC) in the vicinity (from $-3$ to 3 eV) of the Fermi surface. The major contribution to the DOS near Fermi energy comes from Ir-5d states.
\section{The tunable flat band for Sc$_2$Ir$_{4-4p}$Si$_{4p}$}
Here shows the calculated electronic band structures for series compounds from the doping level $x=0.4$ to $x=0.5$. It can be seen that the energy band between $\Gamma$-L (indicated by the red box) becomes flatter and is tuned up to the Fermi surface with the Si doping.
\begin{figure}[H]
\centering
  \setlength{\abovecaptionskip}{-0.1cm}
  \setlength{\belowcaptionskip}{-0.5cm}
  \includegraphics[width=14cm]{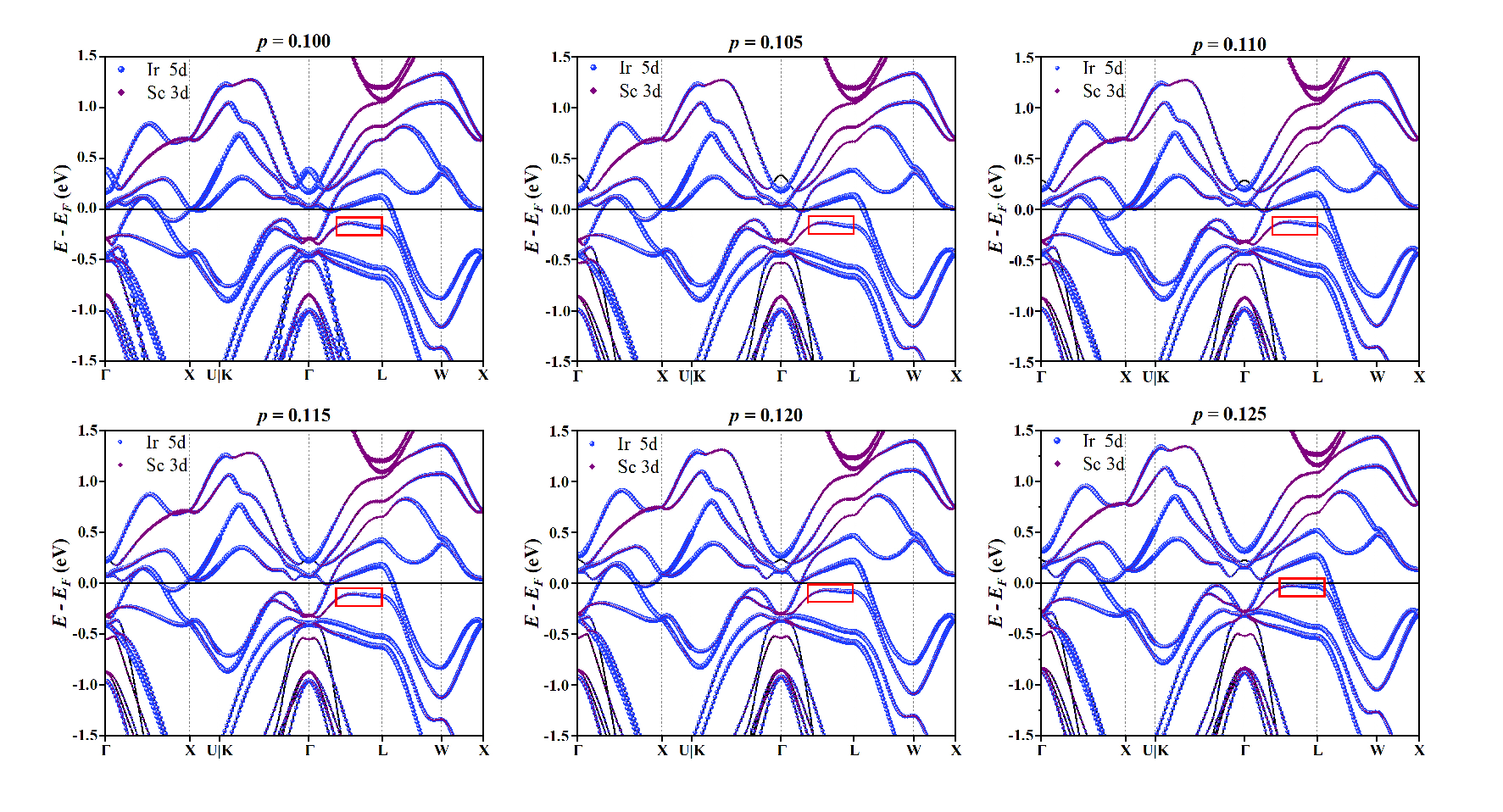}
\caption{The calculated band structures for Sc$_2$Ir$_{4-4p}$Si$_{4p}$ with SOC} \label{figs8}

\end{figure}